# Effect of heavy ion irradiation on microstructure and electron density distribution of zirconium alloy characterised by X-ray diffraction technique

A.Sarkar, P.Mukherjee and P.Barat

**Variable Energy Cyclotron Centre, 1/AF Bidhannagar, Kolkata – 700 064, India.**

## Abstract

Different techniques of the X-ray Diffraction Line Profile Analysis (XRDLPA) have been used to assess the microstructure of the irradiated Zr-1.0%Nb-1.0%Sn-0.1%Fe alloy. The domain size, microstrain, density of dislocation and the stacking fault probabilities of the irradiated alloy have been estimated as a function of dose by the Williamson-Hall Technique, Modified Rietveld Analysis and the Double Voigt Method. A clear signature in the increase in the density of dislocation with the dose of irradiated was revealed. The analysis also estimated the average density of dislocation in the major slip planes after irradiation. For the first time, we have established the changes in the electron density distribution due to irradiation by X-ray diffraction technique. We could estimate the average displacement of the atoms and the lattice strain caused due to irradiation from the changes in the electron density distribution as observed in the contour plots.

## 1. Introduction

Energetic particles such as electrons, heavy ions and neutrons transfer energy to the solid materials primarily by the process of ionization, electronic excitation and also by the displacements of atoms from their original sites [1]. These processes cause a change in the internal microstructure, phase distributions, dimensions and the mechanical properties [2-5] of the target material. Generally neutrons and heavy ions in the MeV range impart so much energy to the primary knock-on that a displacement cascade is produced consisting of highly localized interstitials and vacancies [6] associated with a single initiating event. The process and the reaction pathways by which the displacement and electronic energy are dissipated, determine the structure and the changes of the properties, exhibited by the material. Microstructural evolution in the metals and alloys during the irradiation with the energetic particles has been reviewed thoroughly with an emphasis placed on the underlying defect reaction processes [7]. In case of the light ions such as protons, the damage profiles are much more homogeneous than that of heavy ions. The damage caused by the neutron irradiation is often simulated by using the high-energy particle irradiations, which allow easy variations of the irradiation conditions [6]. By using light or heavy ions, the recoil spectrum can be altered so that it covers the significant ranges of the neutron recoil spectrum [6]. Thus the nature of the radiation damage in the material is affected by the type of ions used for irradiation, alloying elements and the impurity variations [8].

In the present study, we have carried out irradiation with 145 MeV $Ne^{6+}$ with degrader on Zr-1.0%Nb-1.0%Sn-0.1%Fe at different doses. Different techniques of X-ray Diffraction Line Profile Analysis (XRDLPA) have been used to evaluate the effect of the irradiation on the microstructure of the material. XRDLPA is a powerful technique to evaluate the microstructural parameters in a statistical manner [9]. Different techniques of XRDLPA have been widely applied for the evaluation of microstructural parameters in different deformed metals and alloy systems [10, 11]. In our earlier studies we have shown the microstructural variation due to the irradiation with the proton and the oxygen ions on the same alloy [12, 13].

In this work, we have characterized the microstructural parameters by XRDLPA using different model based approaches such as the Williamson-Hall Technique, Modified Rietveld Analysis and the Double Voigt Method on Neon irradiated Zr-1.0%Nb-1.0%Sn-0.1%Fe alloy. The domain size, microstrain, density of dislocation and the stacking fault probabilities of the irradiated alloy have been estimated as a function of dose. For the first time, we have established the changes in the electron density distribution due to irradiation by X-ray diffraction technique on some specific crystallographic planes. We could estimate the average displacements of the atoms and the lattice strain caused due to heavy ion irradiation from the changes of the electron density distribution as observed in the contour plots.

The damage profile as a function of depth from the surface has been characterized in terms of displacements per atom (dpa) for different doses.

## 2. Experimental

Ingot of Zr-1.0%Nb-1.0%Sn-0.1%Fe alloy was prepared in Nuclear Fuel Complex, Hyderabad, India, by double vacuum arc melting technique. It was then β quenched, followed by hot extrusion and cold pilgering for producing fuel cladding tubes of 0.4 mm wall thickness.

Samples of size 10 mm × 10 mm were cut from these tubes and annealed at a temperature of 1023 K for 4 h. The samples were mounted on an aluminum flange and covered with an aluminum foil of thickness 30 μm which was used as a degrader. These samples were then irradiated with 145 MeV $Ne^{6+}$ ions from Variable Energy Cyclotron (VEC), Kolkata, India. Incident energy of the particle on the sample was 110 MeV after degradation. The irradiation doses were $3\times10^{17}$, $8\times10^{17}$, $1\times10^{18}$ and $3\times10^{18}$ $Ne^{6+}$ ions/m$^2$. The flange used for irradiation was cooled by continuous flow of water. During the irradiation, the temperature of the sample did not rise above 313K as measured by the thermocouple connected very close to the sample. The range of the ions in this material and the dpa were obtained by Monte-Carlo simulation technique using the code SRIM 2000 [14].

X-Ray Diffraction (XRD) profiles for each irradiated sample have been recorded from PHILIPS 1710 diffractometer using CuK$_\alpha$ radiation. The range of 2θ was from 25° to 100° and a step scan of 0.02° was used. The time per step was 4 seconds.

## 3. Method of Analysis

In most of the investigations, structural information is extracted from namely the angular positions and the intensities of the Bragg peaks in the diffraction pattern. In the present study, we are interested in the variation of the microstructure and the electron density distribution with irradiation. Generally, the broadening of a Bragg peak arises due to the instrumental broadening, broadening due to the small domain size and the microstrain. However, more detailed information is extractable from the line shapes of the Bragg peaks. The analysis of the line shapes allows one to characterize the microstructure more comprehensively in terms of the mean square microstrain and the average domain size. The Williamson-Hall Technique, Modified Rietveld Method using the whole powder pattern fitting technique and the Double Voigt Analysis have been adopted in the present study in order to analyze the line shapes of the diffraction data of Zr-1.0%Nb-1.0%Sn-0.1%Fe at different doses of irradiation. The instrumental broadening correction was made using a standard defect free Si sample. The electron density distribution and the contour plots were obtained by the Fourier transform of the diffraction patterns.

*Williamson-Hall Technique*

Williamson and Hall [15] assumed that both the size and the strain broadened profiles are Lorentzian. Based on this assumption, a mathematical relation was established between integral breadth ($\beta$), volume weighted average domain size ($D_v$) and the upper value of the microstrain ($\varepsilon$) as follows.

$$\frac{\beta \cos\theta}{\lambda} = \frac{1}{D_v} + 2\varepsilon \left(\frac{2\sin\theta}{\lambda}\right) \quad (1)$$

The plot of $\left(\frac{\beta \cos\theta}{\lambda}\right)$ versus $S = \left(\frac{2\sin\theta}{\lambda}\right)$ gives the value of the microstrain from the slope and the domain size from the ordinate intercept.

*Modified Rietveld Method*

In this method, the diffraction profiles have been modeled by pseudo-Voigt (*pV*) functions using the program LS1 [16].

This program refines simultaneously the lattice parameters, surface weighted average domain size ($D_s$), average microstrain $\langle \varepsilon_L^2 \rangle^{\frac{1}{2}}$ and the preferred orientation parameter P [17, 18] assuming isotropicity in the domain size and the microstrain. The effective domain size ($D_e$) with respect to each fault-affected crystallographic plane was then refined to obtain the best fitting parameters.

XRD peak profiles of Zr-1.0%Nb-1.0%Sn-0.1%Fe show strong crystallographic texture along certain crystallographic directions particularly (002), (101), (102) and (103). The h,k,l values of these planes were incorporated in the program as the preferred oriented planes and the best fit was sought in each case.

Considering the X-ray line profiles to be symmetric in shape, the distributions of the dislocations were assumed to be random. The average density of dislocation (ρ) has been estimated from the relation [19] $\rho = (\rho_D \rho_S)^{\frac{1}{2}}$, where, $\rho_D = \frac{3}{D_s^2}$ (density of dislocation due to domain) and $\rho_S = k \langle \varepsilon_L^2 \rangle / b^2$ (density of dislocation due to strain), k is the material constant and $b$ is the modulus of Burger's vector, $\frac{1}{3}[11\bar{2}0]$. Similarly, $\rho_e$, the density of dislocation at each crystallographic plane has been estimated.

$D_e$ and $D_s$ are related to the deformation stacking fault probability $\alpha$, considering the growth fault probability (β) to be zero in the h.c.p alloys [20-24]. The values of $\alpha$ for various doses are reported.

*Double Voigt Technique*

In this technique, the size and the strain effects are approximated by a Voigt function [25], which is a convolution of Gaussian and Cauchy functions. The equivalent analytical expressions for the Warren-Averbach size-strain separation [26] were then obtained. The Fourier coefficients $F(L)$ in terms of a distance, $L$, perpendicular to the diffracting planes is obtained by the Fourier transform of the Voigt function [25] and can be written as

$$F(L) = \left(-2L\beta_C - \pi L^2 \beta_G^2\right) \qquad (2)$$

where, $\beta_C$ and $\beta_G$ are the Cauchy and Gauss components of the total integral breadth respectively.

$\beta_C$ and $\beta_G$ can be written as:

$$\beta_C = \beta_{SC} + \beta_{DC} \qquad (3)$$

$$\beta_G^2 = \beta_{SG}^2 + \beta_{DG}^2 \qquad (4)$$

where, $\beta_{SC}$ and $\beta_{DC}$ are the Cauchy components of the size and the strain integral breadth respectively and $\beta_{SG}$ and $\beta_{DG}$ are the corresponding Gaussian components.

The size and the distortion coefficients were obtained considering at least two reflections from the same family of crystallographic planes. The surface weighted average domain size $D_s$ and the microstrain $\langle \varepsilon_L^2 \rangle^{\frac{1}{2}}$ are given by the equations:

$$D_s = 1/2\beta_{SC} \qquad (5)$$

$$\langle \varepsilon_L^2 \rangle = \left[\beta_{DG}^2/(2\pi) + \beta_{DC}/(\pi^2 L)\right]/S^2 \text{ where } S = \frac{2\sin\theta}{\lambda} \qquad (6)$$

The volume weighted domain size [27] is given by:

$$D_v = \frac{1}{\beta_S} \text{ where } \beta_S = \frac{\beta\cos\theta}{\lambda}, \text{ integral breadth in the units of } S, (\text{Å})^{-1}.$$

The surface weighted and the volume weighted column-length distribution functions are given by:

$$P_s(L) \propto \frac{d^2 A_S(L)}{dL^2} \qquad (7)$$

$$P_v(L) \propto L \frac{d^2 A_S(L)}{dL^2} \qquad (8)$$

For a size-broadened profile, the size coefficient is given as:

$$A_S(L) = \exp(-2L\beta_{SC} - \pi L^2 \beta_{SG}^2) \qquad (9)$$

From equation (8), we get,

$$\frac{d^2 A_S(L)}{dL^2} = [(2\pi L \beta_{SG}^2 + 2\beta_{SC})^2 - 2\pi \beta_{SG}^2] A_S(L) \qquad (10)$$

Selivanov and Smislov [28] showed that equation (9) is a satisfactory approximation of the surface wieghted column-length distribution function.

### *Estimation of the Electron Density Distribution*

Electron density (ED) $\rho(xyz)$ [29] in a crystal is a periodic function of $\vec{r}$ with periods of the primitive translation vectors in the three crystal axes. Hence, $\rho(xyz)$ can be expanded as

$$\rho(xyz) = \frac{1}{V} \sum_{h,k,l=-\infty}^{\infty} F(hkl) \exp[-2\pi i(hx + ky + lz)] \qquad (11)$$

where $F(hkl)$ is the structure factor for the specific plane $(hkl)$.

We have used the programme GSAS [30] to estimate the electron density function $\rho(xyz)$ and the contour plots for both the unirradiated and irradiated samples.

## 4. Results and discussions

The range of 110MeV $Ne^{6+}$ ion in Zr-1.0%Nb-1.0%Sn-0.1%Fe (obtained by SRIM 2000 calculation) was found to be around 39μm which is of the order of the depth of penetration of $CuK_\alpha$ X-ray in this material. The radiation damage has been assayed by the damage energy deposited causing displacements of atoms. The total target displacements of the collision events calculated by the SRIM 2000 code is shown in Fig. 1. The damage is measured by the dpa. The average dpa for the highest dose sample in Zr-1.0%Nb-1.0%Sn-0.1%Fe was found to be $1.2 \times 10^{-2}$. The average dpa value has been calculated over the total range of 39 μm. The dpa value at 39 μm is found to be 0.13.

Fig. 2 represents a typical XRD profile of the unirradiated and irradiated Zr-1.0%Nb-1.0%Sn-0.1%Fe.

*Williamson-Hall technique*

The Williamson-Hall (WH) technique gives the information of $D_v$ and $\varepsilon$ within the domain from the plot of $\left(\dfrac{\beta \cos\theta}{\lambda}\right)$ against $S$. Fig.3 shows the WH plots for the unirradiated and the irradiated Zr-1.0%Nb-1.0%Sn-0.1%Fe at different doses. For most of the cases, it is seen that $\dfrac{\beta \cos\theta}{\lambda}$ (in $\sin\theta$ scale) shows a linear $S$ dependence. This implies that the shape of the domains is isotropic. It is further observed that the line connecting two orders of (00*l*) type reflections i.e. <002> and <004> and also <101> and <202> yield a considerable slope indicating strong lattice distortion along *<001>* and *<101>* direction respectively. The values of $D_v$ and $\varepsilon$ obtained from the intercept and the slope of WH plots are shown in Table-1. It is observed that $D_v$ decreased by a considerable amount in the first dose of irradiation. The values were found to decrease further with the dose. The value of $\varepsilon$ was found to increase slightly with the dose.

*Modified Rietveld Method*

We have carried out analysis on the XRD patterns of irradiated Zr-1.0%Nb-1.0%Sn-0.1%Fe with the help of modified Rietveld method using program LS1 [16]. The variation of $D_s$, $\langle \varepsilon_L^2 \rangle^{\frac{1}{2}}$ and $\rho$ for these samples have been plotted as a function of dose in Fig.4, Fig.5 and Fig.6 respectively.

Significant changes were found in the values of $D_s$, $\langle \varepsilon_L^2 \rangle^{\frac{1}{2}}$ and $\rho$ with dose in the irradiated samples as compared to the unirradiated one. The values of $D_s$ initially decreased with increasing dose of irradiation but it saturated at higher dose. On the contrary, the average microstrain initially increased with dose of irradiation and consequently saturated at higher dose. The density of dislocation increased significantly for the irradiated samples and the increase was found to be almost an order of magnitude more for the case of the irradiated sample at a dose of $1 \times 10^{18}$ $Ne^{6+}/m^2$ as compared to the unirradiated one. The density of dislocation was also found to saturate with dose.

The reason of the above findings can be explained as follows:

The range of 110 MeV Neon in Zr-1.0%Nb-1.0%Sn-0.1%Fe was found to be 39 $\mu m$ as calculated by SRIM 2000. Neon being a heavy ion, transferred sufficient energy to the primary knock-on atoms and a displacement cascade was produced consisting of highly localized interstitials and vacancies associated with a single initiated event. Moreover, the energy transferred to the lattice atoms was much larger at the end of the trajectory. As the primary recoil proceeded through the sample, loosing energy in successive collisions, the displacement cross-section increased [31]. Thus the distance between the successive displacements decreased and at the end of the track, the recoil collided with practically every atom in its path, creating a very high localised concentration of the vacancies and the interstitials. These mobile point defects interact with the microstructure by long-range diffusion [6]. The main mechanism of the migration of the point migration and their annihilation are based on three reaction paths: (i) the loss of point defects at the extended sinks such as the surfaces, grain boundaries and at the network of the existing dislocations, (ii) the nucleation of the clusters by the homogeneous reactions between the point defects of the same type, (iii) the growth of the defect clusters like the dislocation loops, voids by the agglomeration of the point defects.

For Ne$^{6+}$ ion irradiation, the damage is maximum within a distance of 2-3 μm at the end of the reaction path, where the dpa is 0.13. A concentration gradient of defects in the sample was thus created within a small reaction path of 39μm which helped in the migration of defects. Again, the diffusion coefficient $D_a$ of a particular lattice atom is enhanced due to irradiation [32] and is given by the following equation:

$$D_a = f_v D_v C_v + f_{2v} D_{2v} C_{2v} + f_i D_i C_i + ...... \qquad (12)$$

Thus, $D_a$ is increased by increasing the concentration of different defect species such as the vacancies, di-vacancies, interstitials etc. and also by opening up the other diffusion channels via defect species which are not significantly present in the normal thermally activated diffusion.

In the irradiated sample, the enhancement of radiation induced diffusion is solely responsible for the migration of the vacancies, their agglomeration and the collapsing in the shape of dislocation loops. This is the only mechanism by which density of dislocation is increased in the irradiated samples as Frank-Reed source mechanism for the multiplication of dislocation is absent due to the non-availability of any stress field. The generation of the dislocation by the collapsing of vacancy clusters is only possible when there is an excess vacancy concentration than the equilibrium values. Hence, we could observe an order of magnitude increase in the density of dislocation at a dose of 1x10$^{18}$ Ne$^{6+}$/m$^2$. During irradiation, two competing processes occur simultaneously, one is the generation of vacancies, agglomeration of vacancies and then collapsing into dislocation loops and the other is, their annihilation at the possible sinks. Initially, at the low dose of irradiation (3x10$^{17}$ Ne$^{6+}$/m$^2$), the rate of generation of dislocation loops dominates over the rate of annihilation of the point defects as the sink density is low. So, we found an increase in the density of dislocation. With increasing dose of irradiation, though more vacancies are created, annihilation rate of vacancies also increases as the sink density increases with irradiation. Hence, a saturation was observed in the density of dislocation with the increase in the dose of irradiation.

The size of the domains of the irradiated samples decreased with the increase in the dose of irradiation. The decrease was quite drastic at lower doses and almost saturated at higher doses, as the generation of dislocation did not vary significantly at higher doses.

The effective domain size $D_e$, along different crystallographic directions were also found to decrease with dose as compared to the unirradiated material but the shape of the domains were almost isotropic. We have plotted the projections of $D_e$ (along different directions) on the plane containing the directions <002> and <100>. Only the projections in the first quadrant are shown in Fig.7 for unirradiated and irradiated Zr-1.0%Nb-1.0%Sn-0.1%Fe. It was clearly observed that, $D_e$ was almost isotropic (spherical) with values $<D_e>_{002} \cong 1077$Å and $<D_e>_{100} \cong 1070$Å for the unirradiated sample, $<D_e>_{002} \cong 969$ Å and $<D_e>_{100} \cong 891$ Å at a dose of $3 \times 10^{17}$ Ne$^{6+}$/m$^2$, $<D_e>_{002} \cong 760$Å and $<D_e>_{100} \cong 724$Å at a dose of $8 \times 10^{17}$ Ne$^{6+}$/m$^2$, $<D_e>_{002} \cong 417$Å and $<D_e>_{100} \cong 419$Å at a dose of $1 \times 10^{18}$ Ne$^{6+}$/m$^2$ and $<D_e>_{002} = 372$ Å and $<D_e>_{100} = 370$ Å at a dose of $3 \times 10^{18}$ Ne$^{6+}$/m$^2$. These values clearly signify that the shape of the domains did not change with dose though the variations in the size of the domains were significant with the increasing dose as compared to the unirradiated sample.

This analysis also revealed that the density of dislocation at each crystallographic plane has increased as a function of dose, as shown in Table-2. The analysis is important as the estimated values of the density of dislocation on various planes particularly on the slip planes provide the information about the flow property of any material. The flow stress of metals is proportional to the square root of the density of dislocation [33]. In order to predict the number and the nature of the active slip, systems it is therefore essential to know the density of dislocation per slip plane [34].

The microstrain values at $L$=50Å along different crystallographic directions for the alloy at different doses are shown in Table-2. The values showed an increasing trend for irradiated samples as compared to the unirradiated one.

The deformation fault ($\alpha$) was found to be negligibly small for the unirradiated and the irradiated samples.

*Double Voigt Analysis*

The general conclusions obtained from the simple WH plot can be further substantiated by a detailed analysis. In this analysis, both the size and the strain broadened profiles were approximated by a Voigt function. The Cauchy and the Gaussian components ($\beta_{SC}, \beta_{SG}, \beta_{DC}$ and $\beta_{DG}$) of the size and the strain broadened profiles were then separated along <001> and listed in Table-3. From Table-3, it is observed that, in general size broadened profiles had both Cauchy and Gaussian components of the integral breadths.

The volume weighted column-length distribution function $P_v(L)$ along <00$l$> normal to the diffraction plane (00$l$) has been shown in Fig.8. It is clear that the column length distribution is much wider for the unirradiated sample and it is found to narrow down with the increasing dose of irradiation.

Using different model based approaches of XRDLPA techniques, the microstructure of the irradiated Zr-1.0%Nb-1.0%Sn-0.1%Fe at different doses have been characterized. All these techniques are based on the profile shape and the broadening of the diffraction peak. These techniques have limitations in characterising the small defects particularly small interstitial clusters which do not cause broadening of the peak but contribute to the background values close to the Bragg peak [35]. Scattering of X-rays from interstitial clusters [36] are diffuse scattering very close to the Bragg peak (Huang Scattering). Thus, the complete information of the microstructure of the irradiated samples can be obtained from the X-ray diffraction techniques by the combined studies of the diffraction pattern in the Bragg peak region (coherent scattering) and in the background region (diffuse scattering close to the Bragg peak). As in our case, the experiments were carried out at room temperature, the diffuse scattering near the Bragg peak region due to small interstitial clustering are superimposed by thermal diffusion scattering. Hence, the line profile analysis could characterise only those microstructurral parameters which are responsible for the broadening of the diffraction peaks.

*Electron Density Distribution*

The contour plots of the electron density (ED) for the unirradiated samples and the sample irradiated at a dose of $3\times10^{17}$ $Ne^{6+}/m^2$, $8\times10^{17}$ $Ne^{6+}/m^2$ and $3\times10^{18}$ $Ne^{6+}/m^2$ have been shown in Fig. 9 and compared with the calculated one. Each contour plot is 10Å in size on (0001) plane at z=0, z denotes the perpendicular distance along c. In order to calculate the lattice parameter 'a' from these contour plots, the image of the each contour plot was analyzed using MATLAB and the center of the position of the atoms was accurately determined. The estimated values of 'a' for the calculated and the irradiated samples at doses $3\times10^{17}$ $Ne^{6+}/m^2$, $8\times10^{17}$ $Ne^{6+}/m^2$, $1\times10^{18}$ $Ne^{6+}/m^2$ and $3\times10^{18}$ $Ne^{6+}/m^2$ were 3.205 Å, 3.213 Å, 3.214 Å, 3.216 Å and 3.218 Å respectively. Moreover we found that the c/a values (obtained from GSAS) decreased gradually with dose. So, it is obvious that there was an expansion of 'a' with increasing dose and the calculated lattice strain was found to be of the order of $10^{-3}$ along $\langle a \rangle$. As a result, there was a change in the average spatial charge distribution which is clearly evident in Fig. 9. The lattice strain values obtained from the contour plot corroborate with the values of the microstrain estimated from the line profile analysis of the diffraction peaks. In Fig. 10 we have drawn a contour map of the unirradiated sample on the plane (0001) at z=0.25. This figure of the unirradiated sample shows the effect on ED for the atom at z=0.25, which is shown by the dense circular contours. The other prominent contours on the same map of almost circular shape are due to the presence of the atom at z=0.75. We have compared this contour plot with the observed ED plots at different doses. It is clear that at low dose i.e. $3 \times 10^{17} Ne^{6+}/m^2$, the ED plot did not change significantly. On the contrary a significant change was observed at higher doses where the contours for the atom at z=0.25 on the ED plot have been distorted. The values of charge density due to the presence of atom at z=0.75 decreased with dose and were found to be negative at higher doses. This observation clearly signifies the shifting of atom at z=0.75 from its position and as a result, the contour plot for the atom at z=0.25 also gets perturbed. The electron density of the outermost contour for the atom at z=0.25 for the each plot has been compared and was found to vary from 1.21 e/Å$^3$ for the lowest dose sample to 5.07 e/Å$^3$ for the highest dose sample. This depicts that there was a localization of charge density with the increase in dose. Moreover, we see that the variation of the maximum electron density, $\rho_{max}$ with z was found to follow the same pattern for the unirradiated and the irradiated samples as shown in Fig. 11. This may be attributed to the fact that the irradiation has not caused any amorphisation or change in phase, only an average displacement of atoms from their lattice position has occurred.

## 5. Conclusion

XRDLPA can be used as a technique to analyze the change in the microstructure of the materials due to radiation damage. In this work, the microstructure of the unirradiated and the irradiated Zr-1.0%Nb-1.0%Sn-0.1%Fe has been reliably assessed by XRDLPA using different model based approaches. The microstructural parameters like average and effective domain sizes, microstrain within the domains have been characterised as a function of dose. The density of dislocation and the stacking fault probability have been estimated from these values. The analysis revealed that there was a significant decrease of surface weighted average domain size ($D_s$) with dose. The damage associated with neon beam (being heavy ion) was quite extensive and produced highly localized concentration of defects, particularly vacancies and interstitials. These vacancy clusters then collapsed in the shape of dislocation loops and the density of dislocation increased accordingly. This analysis also estimated the average density of dislocation in the major slip planes. The deformation (stacking) fault probability was found to negligible for this alloy even with increasing dose of irradiation. The column length distribution was found to be narrower at highest dose of irradiation for this alloy. For the first time we have established the changes in the electron density distribution due to irradiation by X-ray diffraction technique. The contour plots of ED of the irradiated samples were found to change significantly as compared to the unirradiated one. The plots clearly depicted an average displacement of atoms from their lattice positions and also the localization of charge density with increasing dose of irradiation.

**Table-1: Results of Williamson-Hall Plot**

| Sample | Dose | Volume weighted average domain size ($D_v$) (Å) (±10%) | Microstrain ($\varepsilon$) ($10^{-3}$) (±5%) |
|---|---|---|---|
| Zirlo | Unirradiated | 1000 | 1.8 |
| | $3 \times 10^{17}$ Ne$^{6+}$/m$^2$ | 500 | 1.8 |
| | $8 \times 10^{17}$ Ne$^{6+}$/m$^2$ | 512 | 1.9 |
| | $1 \times 10^{18}$ Ne$^{6+}$/m$^2$ | 454 | 2.0 |
| | $3 \times 10^{18}$ Ne$^{6+}$/m$^2$ | 385 | 2.2 |

**Table-2: Microstrain and density of dislocation for Zr-1.0%Nb-1.0%Sn-0.1%Fe at different doses**

| Sample | Dose (Ne$^{6+}$/m$^2$) → | Microstrain ($10^{-3}$) Max. error ±0.00005 | | | | | Dislocation density ($10^{15}$)(m$^{-2}$) Max. error ±($6\times10^{13}$) | | | | |
|---|---|---|---|---|---|---|---|---|---|---|---|
| | | Unir. | $3\times 10^{17}$ | $8\times 10^{17}$ | $1\times 10^{18}$ | $3\times 10^{18}$ | Unir. | $3\times 10^{17}$ | $8\times 10^{17}$ | $1\times 10^{18}$ | $3\times 10^{18}$ |
| Zr-1.0%Nb-1.0%Sn-0.1%Fe | Fault un-affected | | | | | | | | | | |
| | 002 | 0.6 | 1.4 | 1.5 | 1.1 | 1.6 | 0.1 | 0.3 | 0.5 | 0.6 | 1.0 |
| | 004 | 0.6 | 1.4 | 1.5 | 1.1 | 1.6 | 0.1 | 0.3 | 0.5 | 0.6 | 1.0 |
| | 100 | 0.7 | 2.3 | 2.6 | 1.9 | 2.1 | 0.2 | 0.6 | 0.9 | 1.0 | 1.3 |
| | 110 | 0.7 | 1.4 | 2.0 | 1.4 | 1.8 | 0.1 | 0.5 | 0.6 | 0.8 | 1.1 |
| | 112 | 0.7 | 1.0 | 1.7 | 1.2 | 1.7 | 0.2 | 0.4 | 0.5 | 0.7 | 1.0 |
| | **Fault affected** | | | | | | | | | | |
| | **101** | 0.6 | 1.3 | 2.1 | 1.6 | 1.9 | 0.2 | 0.5 | 0.7 | 0.9 | 1.2 |
| | **102** | 0.7 | 1.2 | 1.8 | 1.3 | 1.7 | 0.1 | 0.4 | 0.6 | 0.7 | 1.0 |
| | **103** | 0.7 | 1.2 | 1.7 | 1.2 | 1.7 | 0.2 | 0.6 | 0.5 | 0.7 | 1.0 |
| | **104** | 0.7 | 1.2 | 1.6 | 1.2 | 1.6 | 0.1 | 0.4 | 0.5 | 0.7 | 1.0 |

| Stacking fault probability ($\alpha$) ($10^{-4}$) | | | | |
|---|---|---|---|---|
| -0.20 | -12.4 | 0.09 | 0.05 | 0.03 |

**Table-3: Results of Double Voigt Method for Zr-1.0%Nb-1.0%Sn-0.1%Fe at different doses**

| Sample | Dose | $\beta_{SC}$ $(10^{-3})$ | $\beta_{SG}$ $(10^{-3})$ | $\beta_{DC}$ $(10^{-3})$ | $\beta_{DG}$ $(10^{-3})$ | $D_S$ (Å) | $\varepsilon$ $(10^{-3})$ | $D_V$ (Å) |
|---|---|---|---|---|---|---|---|---|
| *Zirlo* [001] | Unirradiated | 0.03 | 0.11 | 0.05 | 0.10 | 609 | 0.52 | 723 |
| | 3x10$^{17}$ Ne$^{6+}$/m$^2$ | 0.98 | 1.60 | 0.0 | -- | 376 | -- | 438 |
| | 8x10$^{17}$ Ne$^{6+}$/m$^2$ | 0.0 | 2.49 | 0.76 | -- | 321 | 1.16 | 402 |
| | 1x10$^{18}$ Ne$^{6+}$/m$^2$ | 0.57 | 2.26 | 0.59 | -- | 316 | 1.46 | 378 |
| | 3x10$^{18}$ Ne$^{6+}$/m$^2$ | 0.40 | 2.93 | 0.77 | -- | 257 | 1.82 | 314 |

**Figure Captions**

Fig. 1. Damage profile of 110 MeV Ne$^{6+}$ in Zr-1.0%Nb-1.0%Sn-0.1%Fe.

Fig. 2. Typical XRD patterns of the unirradiated and irradiated Zr-1.0%Nb-1.0%Sn-0.1%Fe.

Fig. 3. Williamson-Hall plots for the unirradiated and irradiated Zr-1.0%Nb-1.0%Sn-0.1%Fe at different doses.

Fig. 4. Variation of average domain size as function of dose.

Fig. 5. Variation of average microstrain as function of dose.

Fig. 6. Variation of average density of dislocation as function of dose.

Fig. 7. Projections of effective domain size on the plane containing the directions <002> and <100> (First quadrant) for unirradiated and irradiated Zr-1.0%Nb-1.0%Sn-0.1%Fe at different doses.

Fig. 8. Volume weighted column length distribution function for unirradiated and irradiated Zr-1.0%Nb-1.0%Sn-0.1%Fe at different doses.

Fig. 9. Electron density distribution maps on (0001) plane at z=0 for Zr-1.0%Nb-1.0%Sn-0.1%Fe (a) unirradiated and irradiated at (b) 3×10$^{17}$ Ne$^{6+}$/m$^2$ (c) 8×10$^{17}$ Ne$^{6+}$/m$^2$ (d) 3×10$^{18}$ Ne$^{6+}$/m$^2$ doses.

Fig. 10. Electron density distribution maps on (0001) plane at z=0.25 for Zr-1.0%Nb-1.0%Sn-0.1%Fe (a) unirradiated and irradiated at (b) 3×10$^{17}$ Ne$^{6+}$/m$^2$ (c) 8×10$^{17}$ Ne$^{6+}$/m$^2$ (d) 1×10$^{18}$ Ne$^{6+}$/m$^2$ (e) 3×10$^{18}$ Ne$^{6+}$/m$^2$ doses.

Fig. 11. Variation of maximum electron density with z at different doses.

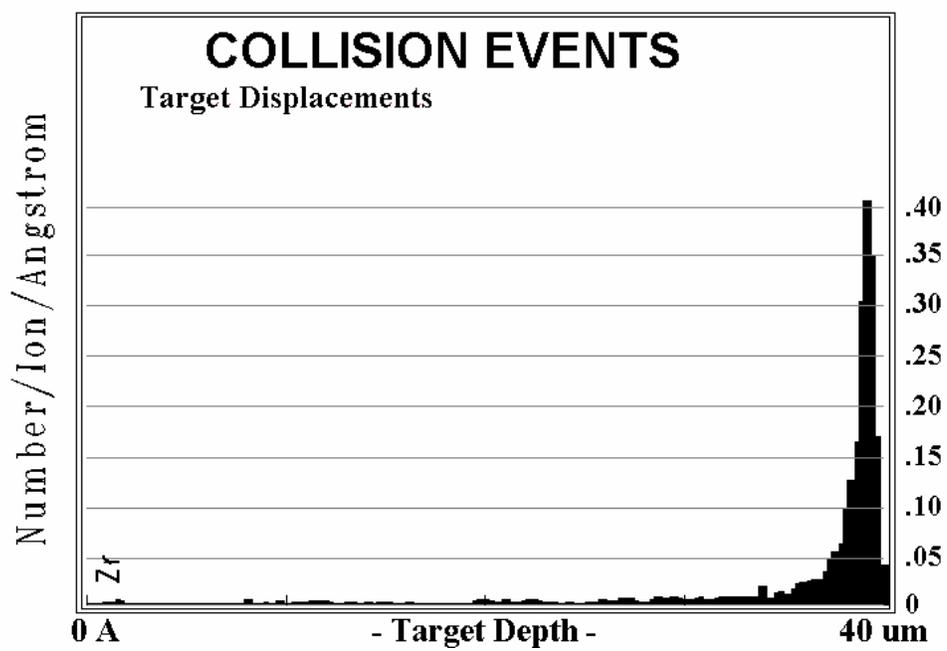

Fig. 1

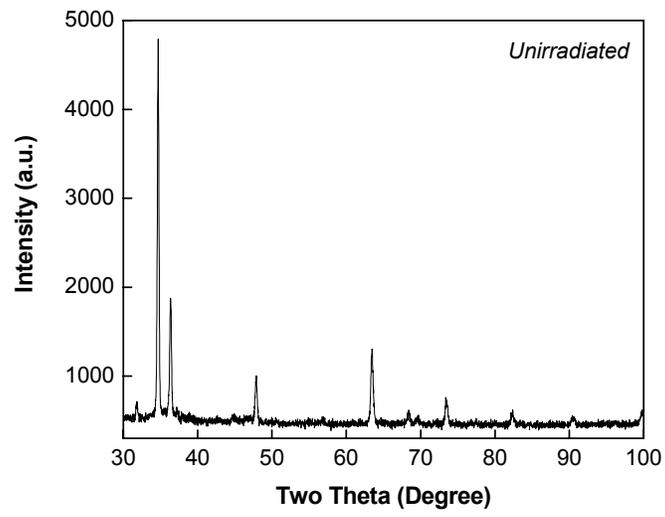

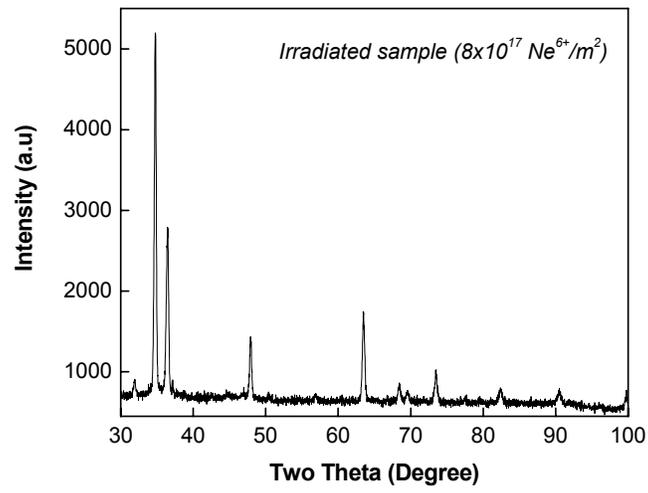

**Fig. 2**

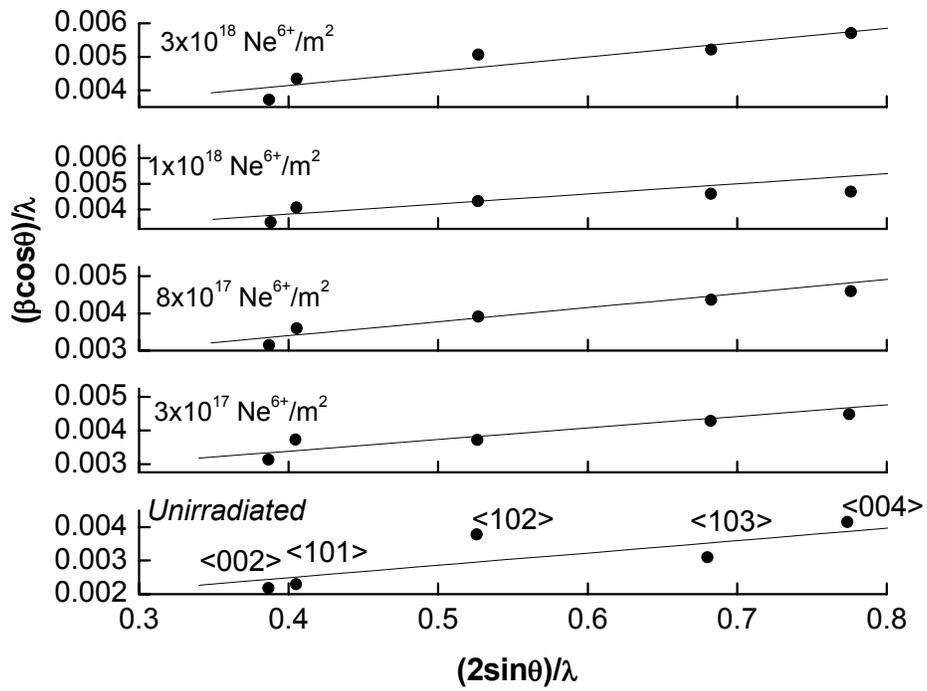

**Fig. 3**

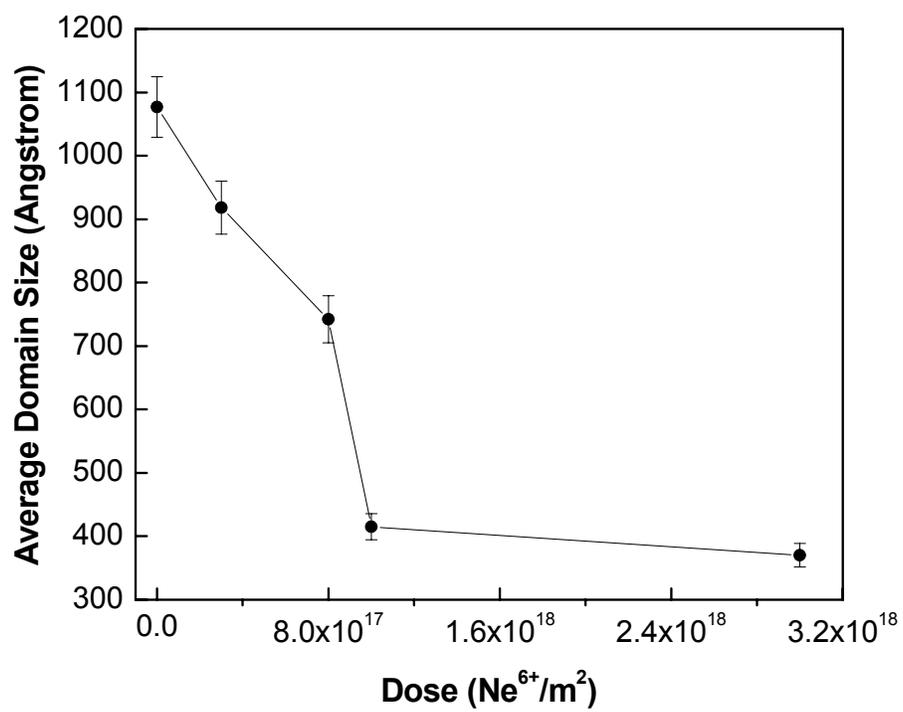

**Fig. 4**

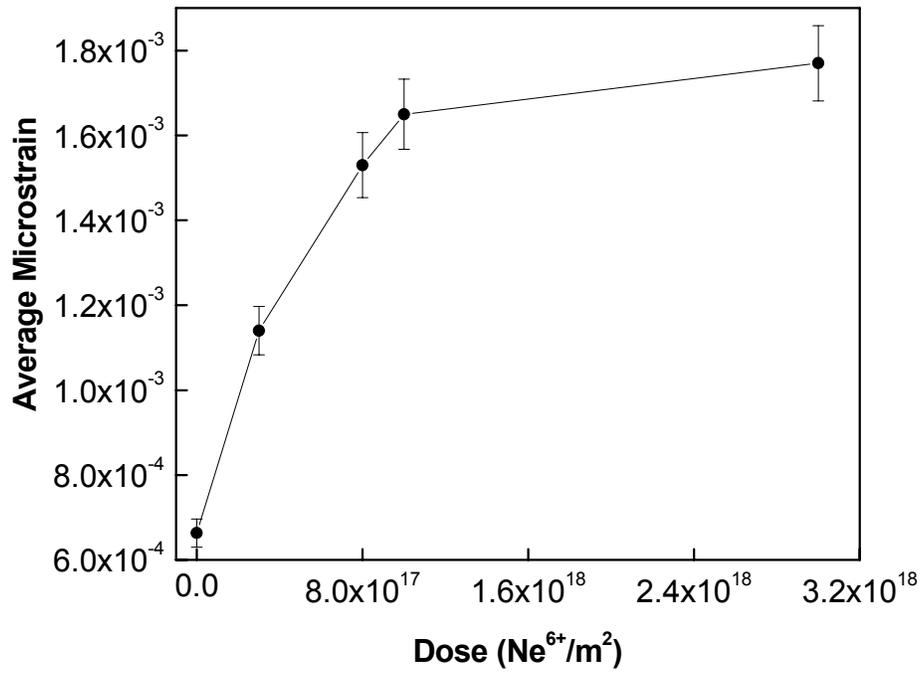

**Fig. 5**

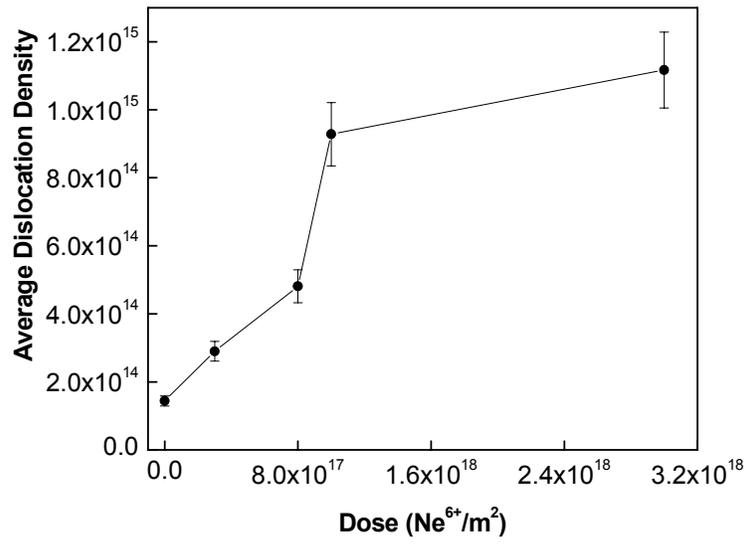

**Fig. 6**

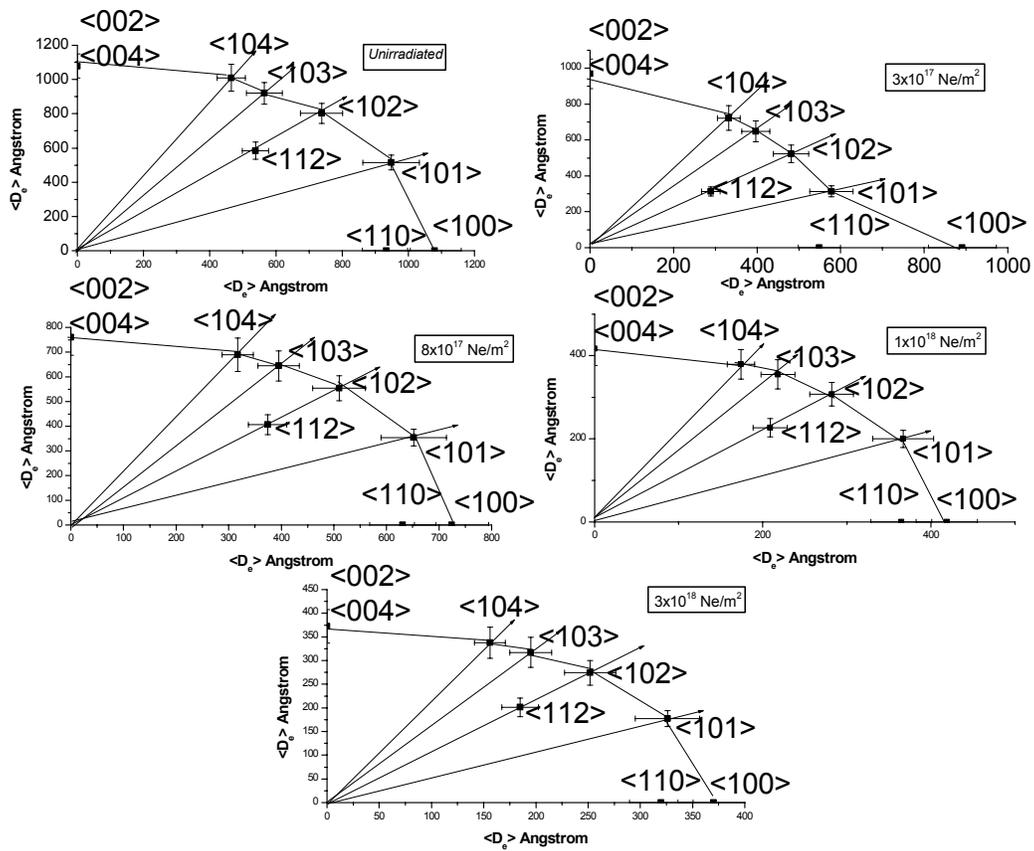

**Fig. 7**

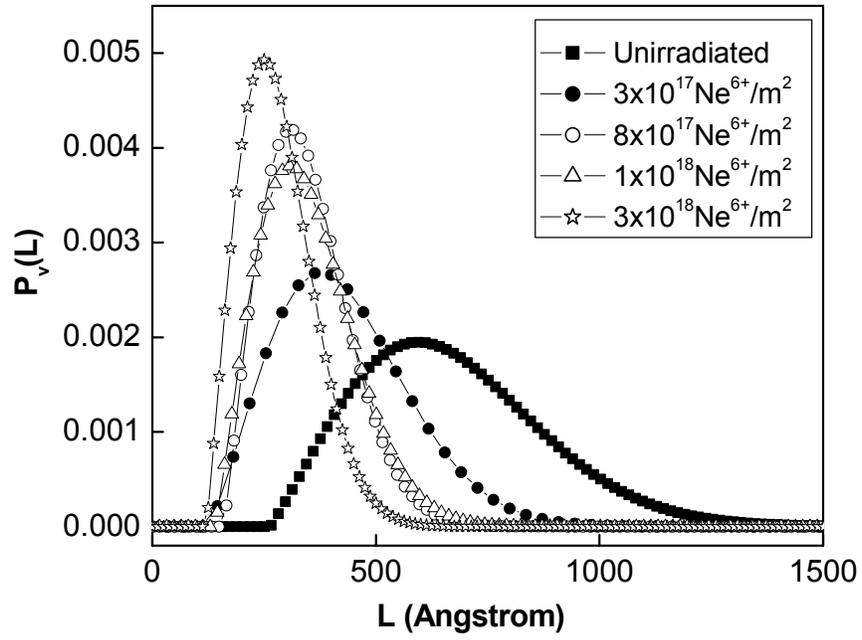

**Fig. 8**

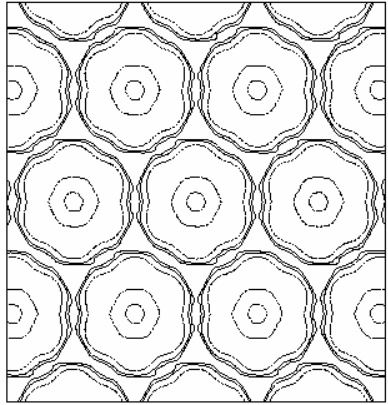 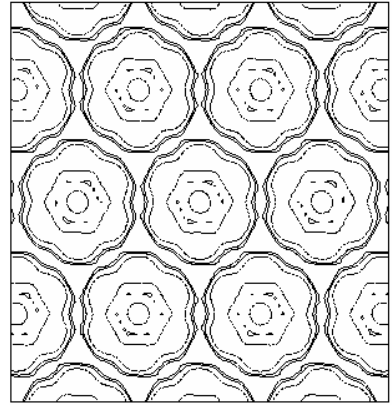
(a)　　　　　　　　　　(b)

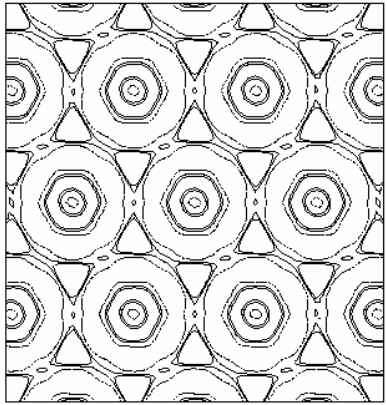 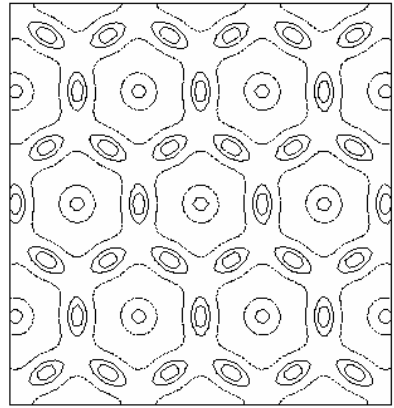
(c)　　　　　　　　　　(d)

**Fig. 9**

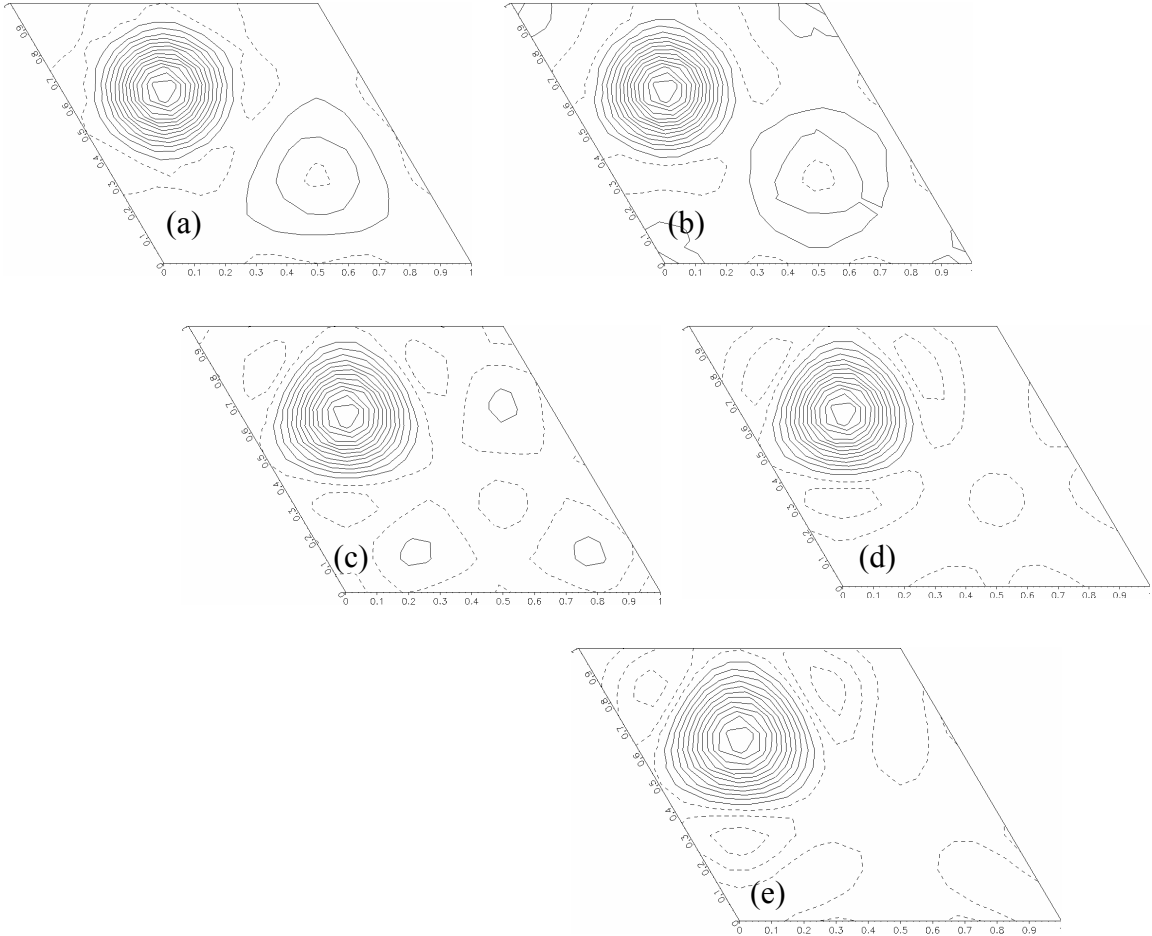

**Fig. 10**

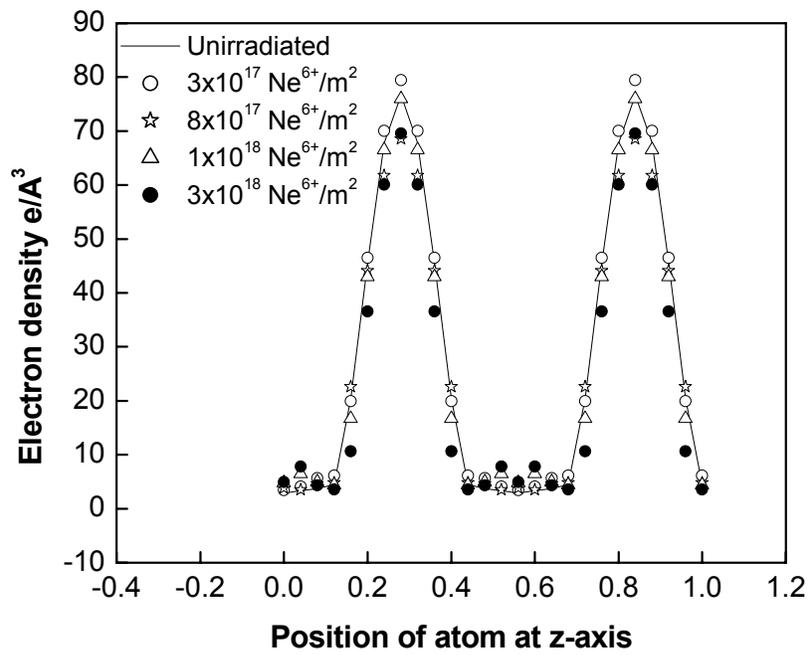

**Fig. 11**